\documentclass[english]{scrartcl}
\usepackage[T1]{fontenc}
\usepackage[latin1]{inputenc}
\usepackage{graphicx}

\makeatletter

 \newcommand{\lyxaddress}[1]{
   \par {\raggedright #1 
   \vspace{1.4em}
   \noindent\par}
 }

\usepackage{babel}
\makeatother
\begin{document}

\title{Random trading market: Drawbacks and a realistic modification}

\author{Srutarshi Pradhan}

\maketitle

\lyxaddress{\begin{center}Department of Physics, NTNU, Trondheim, Norway\end{center}}

\begin{abstract}
We point out some major drawbacks in random trading market models
and propose a realistic modification which overcomes such drawbacks
through `sensible trading'. We apply such trading policy in different
situations: a) Agents with zero saving factor b) with constant saving
factor and c) with random saving factor --in all the cases the richer
agents seem to follow power laws in terms of their wealth (money)
distribution which support Pareto's observation. 
\end{abstract}

\section*{Introduction }

Pareto power law \cite{Pareto} in wealth distribution has become
a hot topic nowadays. Since last decade physicists are putting a lot
of efforts to study economic market through suitable models \cite{Basics,monopoly,analytic,recent}.
They consider economic market as a multi-agent interacting system
and try to analyze the market through known tools of statistical physics
although in reality the economic market is much more complex and the
agents are inherently different from each other--so cannot be hoped
to behave similarly. To start with, a closed economy market has been
considered having some money exchange interaction --the main intention
is to find out the distribution of wealth (money) among the agents
and to search for a suitable exchange interaction which can produce
Pareto like power law distribution of wealth.

\section*{Random trading market}

The molecules in an ideal gas interact freely-- and the kinetic theory
finds the energy distribution among the molecules. If the agents of
model market exchange their money through such free interactions --that
market is called a random trading market \cite{Basics}. Two types
of basic random trading are possible:

\subsection*{Type-I trading}

Two interacting agents ($i$ and $j$) put all their money ($m_{i}$
and $m_{j}$) together. Then one agent takes a random part of the
total money and rest money goes to other agent.

\begin{center}$total=m_{i}+m_{j}=m_{i}^{\prime}+m_{j}^{\prime}$ \end{center}

\begin{center}$m_{i}^{\prime}=\epsilon\times total$; $m_{j}^{\prime}=total-m_{i}^{\prime}$\end{center}

\begin{center}with $0\leq\epsilon\leq1$\end{center}

Here agents do not save anything and put all their money for trading.
This type of trading rule results \cite{Basics} Gibbs distribution
of the money distribution in the market at steady state having the
form:

\begin{center}$P(m)\sim e^{-m/T}$\end{center}

\noindent where $T$ is the average money (total money/total agent)
of the market. Clearly this distribution is similar to the energy
distribution within the molecules of ideal gas. If the agents do save
some portions (fixed or random) of their money then the total money
available for random trading gets reduced. We can write the general
money exchange rule for two agents having money $m_{i}$ and $m_{j}$
and saving factors $s_{i}$ and $s_{j}$ respectively:

\begin{center}$total=m_{i}+m_{j}$\end{center}

\noindent \begin{center}$(tot)_{av}=m_{i}(1-s_{i})+m_{j}(1-s_{j})$\end{center}

\begin{center}$m_{i}^{\prime}=(m_{i}\times s_{i})+\epsilon\times(tot)_{av}$;
$m_{j}^{\prime}=total-m_{i}^{\prime}$\end{center}

For fixed saving factor ($s_{i}=s_{j}=s$) we get a most probable
type money distribution \cite{Basics,analytic} and for random saving
factor ($0\leq s_{i}\leq1$) the money distribution shows \cite{recent}
Pareto power law with exponent $-2$.

\subsection*{Type-II trading}

Another type of trading may occur \cite{monopoly} when two interacting
agents put same amount of money for random exchange i.e, the richer
agent put an amount just equal to that of the poor agent. Then as
before one agent takes a random part of the total money (available
for trading) and rest money goes to other agent.

\begin{center}$total=m_{i}+m_{j}=m_{i}^{\prime}+m_{j}^{\prime}$\end{center}

If $m_{i}<m_{j}$

\begin{center}$(tot)_{av}=2\times m_{i}$\end{center}

\begin{center}$m_{i}^{\prime}=\epsilon\times(tot)_{av}$; $m_{j}^{\prime}=total-m_{i}^{\prime}$\end{center}

If $m_{i}\geq m_{j}$

\begin{center}$(tot)_{av}=2\times m_{j}$\end{center}

\begin{center}$m_{j}^{\prime}=\epsilon\times(tot)_{av}$; $m_{i}^{\prime}=total-m_{j}^{\prime}$\end{center}

We can generalize the above scheme for two agents ($i$ and $j$)
with saving factors $s_{i}$ and $s_{j}$ as 

\noindent \begin{center}$total=m_{i}+m_{j}$;\end{center}

If $m_{i}<m_{j}$

\begin{center}$(tot)_{av}=2\times m_{i}(1-s_{i})$\end{center}

\begin{center}$m_{i}^{\prime}=(m_{i}\times s_{i})+\epsilon\times(tot)_{av}$;
$m_{j}^{\prime}=total-m_{i}^{\prime}$\end{center}

If $m_{i}\geq m_{j}$

\begin{center}$(tot)_{av}=2\times m_{j}(1-s_{j})$\end{center}

\begin{center}$m_{j}^{\prime}=(m_{j}\times s_{j})+\epsilon\times(tot)_{av}$;
$m_{i}^{\prime}=total-m_{j}^{\prime}$\end{center}

\section*{Major drawbacks}

\subsection*{In Type-I trading}

1) The random exchange in Type-I trading market creates `\textbf{\emph{insecurity}}'
problem. Here richer agents put more money (beyond saving) to trade
with poor agents --even with agents having no money. Therefore the
richer agent always finds greater probability to loose than to gain
from an exchange interaction. It may happen that the richer agent
looses all his money in one interaction. Thus Type-I trading favors
the poor agents and it is really a nightmare to the richer agents. 

\vskip.2in

\noindent 2) Type-I trading shows Pareto power law when the agents
have random saving factor drawn from an interval $0\leq s_{i}\leq1$.
But there is one important restriction that the interval has to includes
the value $1$ (or very nearly $1$). If we take an interval $0\leq s_{i}\leq0.8$
-there will not be any robust power law --that means some agents with
very high saving factor have to be present in the market who always
gain money from the interactions and do not loose. Therefore to achieve
Pareto power law this model pre-assigned some agents as permanent
gainer --which weakens the model itself.

\subsection*{In Type-II trading}

Although Type-II trading seems realistic, it has a basic problem that
it gradually tends toward `\textbf{\emph{monopoly}}' market where
all money goes to a single agent making all others simply beggar.
This happens because when one agent losses all his money he cannot
take part in further money exchange as he does not afford some money
for trade. Thus the agents having `zero' money remain `outcast' from
the society and their number increases as the interactions go on.

\section*{The `sensible trading' scheme}

We propose a \textbf{`sensible trading'} scheme among the agents to
avoid the aforesaid drawbacks. This is a mixture of Type-I and Type-II
trading: Interactions follow Type-II trading with a probability $p$
and obey Type-I trading with probability ($1-p$). Thus the general
money exchange scheme for this `sensible' trading appears as:

\noindent \begin{center}$total=m_{i}+m_{j}=m_{i}^{\prime}+m_{j}^{\prime}$\end{center}

\noindent For $r\leq p$

\noindent \begin{center}$(tot)_{av}=m_{i}(1-s_{i})+m_{j}(1-s_{j})$\end{center}

\begin{center}$m_{i}^{\prime}=(m_{i}\times s_{i})+\epsilon\times(tot)_{av}$;
$m_{j}^{\prime}=total-m_{i}^{\prime}$\end{center}

\noindent For $r>p$

If $m_{i}<m_{j}$

\begin{center}$(tot)_{av}=2\times m_{i}(1-s_{i})$\end{center}

\begin{center}$m_{i}^{\prime}=(m_{i}\times s_{i})+\epsilon\times(tot)_{av}$;
$m_{j}^{\prime}=total-m_{i}^{\prime}$\end{center}

If $m_{i}\geq m_{j}$

\begin{center}$(tot)_{av}=2\times m_{j}(1-s_{j})$\end{center}

\begin{center}$m_{j}^{\prime}=(m_{j}\times s_{j})+\epsilon\times(tot)_{av}$;
$m_{i}^{\prime}=total-m_{j}^{\prime}$\end{center}

Here $r$ is a random number uniformly distributed between $0$ and
$1$. Clearly when $p=0$ only Type-I trading is possible and when
$p=1$ we will have Type-II trading only. Therefore we can reproduce
Type-I and Type-II tradings correctly from this generalized `sensible
trading' scheme. The behavior of this scheme becomes interesting when
$p$ value lies in between the above extremes --so that both --Type-I
and Type-II tradings play their role. We call it a `sensible market'
where Type-II trading dominates much over Type-I trading because Type-II
has no extra risk.  We observe that for $p>0.9$ -the market seems
to approach toward `monopoly trend' and for $p\leq0.5$ Type-I trading
dominates resulting Gibbs-like free-market. Therefore we keep $p$
values in the range $0.5<p\leq0.9$ to have a `sensible market'. 

\vskip.2in

Let us discuss how does this scheme overcome the drawbacks:

\vskip.1in

1) Here the richer agents feel secured as in most of the cases they
apt Type-II trading with poor agents and there is no risk to loose
more. 

2) If some agents loose all their money --still they can interact
with others through Type-I trading (with $1-p$ probability) and can
gain money to become richer. This resists the `\textbf{\emph{monopoly}}'
trend and drives the system toward a steady state. 

\vskip.3in

Now we are going to find (numerically) the money distributions following
`\textbf{sensible trading}' scheme in different situations. We choose
two $p$ values: $p=0.8$ and $p=0.9$ to make the market more sensitive.
Also we choose number of agents $N=100$ and the average money $T=100$
and number of interactions $=5000000$ in each case.

\subsection*{Agents have no saving factor }

\begin{center}\includegraphics[%
  width=3in,
  height=3in]{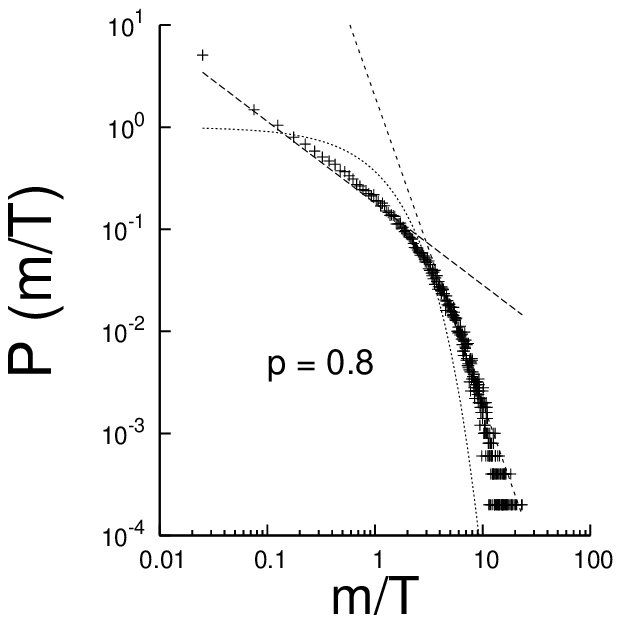}\end{center}

\vskip.2in

\textbf{\small Fig. 1:} {\small The dotted straight lines represents
power laws with exponents $-0.8$ and $-4$ respectively. The curved
dotted line is the plot of $exp(-m/T)$. Averages are taken over $4000$
samples.}{\small \par}

\vskip.2in

Without any saving factor the `sensible' market shows two power laws
and demands that the poor agents and the richer agents obey different
power law behavior. The deviation of the distribution function from
the free market (Gibbs law) is prominent.

\newpage

\subsection*{Agents have constant saving factor ($s$)}

\includegraphics[%
  width=2.5in,
  height=2.5in]{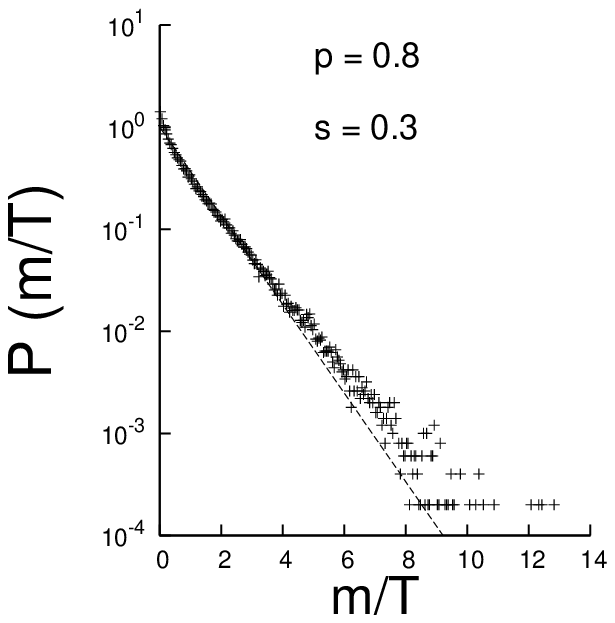}\hskip.1in \includegraphics[%
  width=2.5in,
  height=2.5in]{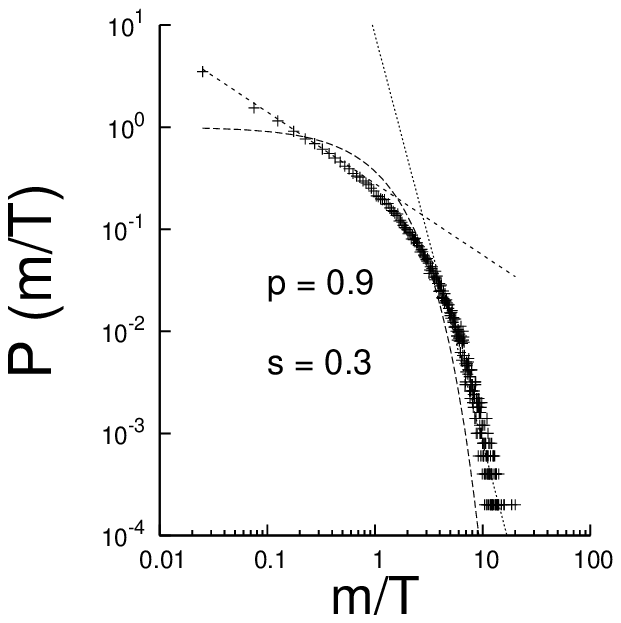}

\begin{center}\includegraphics[%
  width=2.5in,
  height=2.5in]{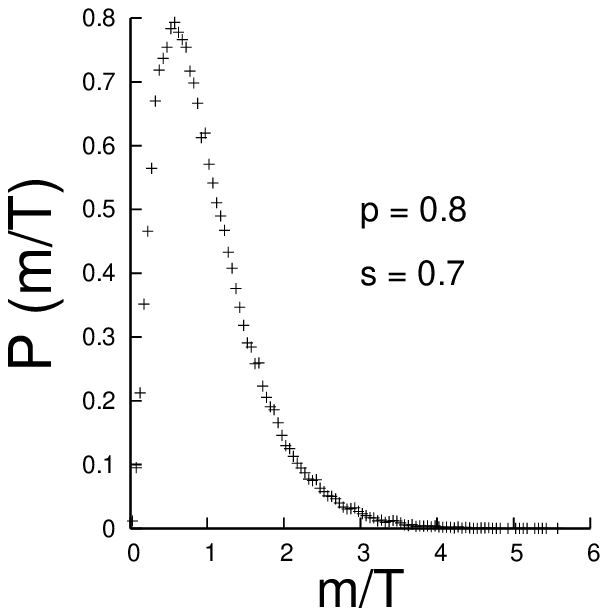}\end{center}

\vskip.2in

\textbf{\small Fig. 2:} {\small The dotted straight line in first
plot represents $exp(-m/T)$ and in second plot presents power laws
with exponents $-0.7$ and $-4$ respectively. Averages are taken
over $1000$ samples.}{\small \par}

\vskip.2in

In Gibbs like free market constant saving factor results most probable
type distribution. But in the sensible market we observe that for
low saving factor the distributions almost follow exponential laws
and this exponential behavior deviates (some power laws appear) as
we increase $p$ values. On the other hand for high saving factor
the most probable type distributions appear. 

\newpage

\subsection*{Agents have random saving factor ($s_{i}$)}

\begin{center}\includegraphics[%
  width=2.5in,
  height=2.5in]{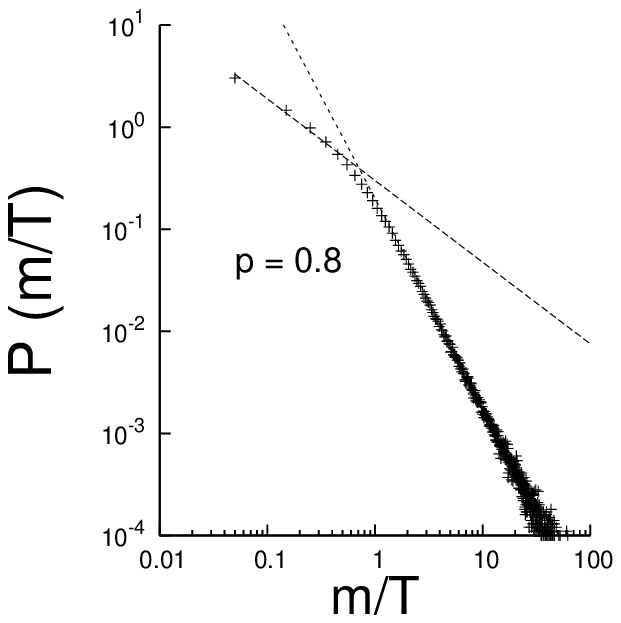}\end{center}

\includegraphics[%
  width=2.5in,
  height=2.5in]{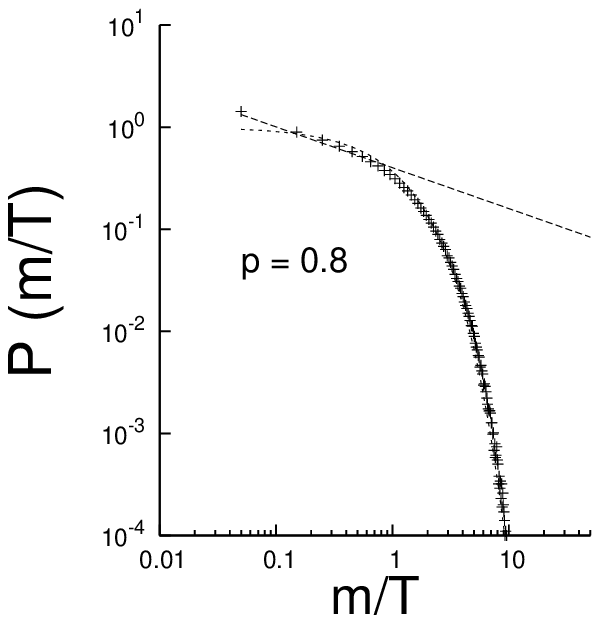}\hskip.2in\includegraphics[%
  width=2.5in,
  height=2.5in]{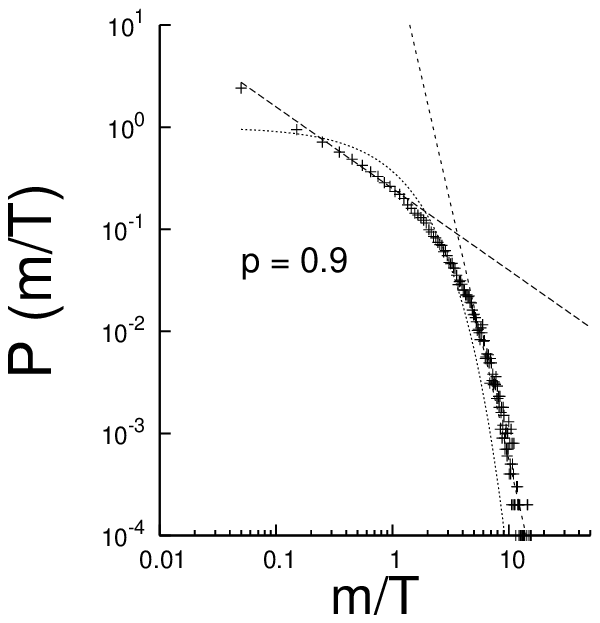}

\vskip.2in

\textbf{\small Fig. 3:} {\small In the first plot we take $0\leq s_{i}\leq1$
and in second and third plot that range has been reduced to $0\leq s_{i}\leq0.8$.
The dotted lines represent power laws having different exponents:
in first plot $-0.8$ and $-2$, in second plot $-0.4$ and in third
plot $-0.8$ and $-5$. Averages are taken over $1000$ samples.}{\small \par}

\vskip.2in

For random saving case we find two distinct power laws when the random
factor is chosen from the interval $0\leq s_{i}\leq1$. Also the market
shows power law distribution with reduced range of random factor ($0\leq s_{i}\leq0.8$)
as $p$ value increases.

\newpage

\section*{Conclusions}

A free market (Type-I trading) basically runs through \emph{`gambling'}
and a restricted market (Type-II trading) gradually becomes a `\emph{monopoly}'
market. But a careful mixing of Type-I and Type-II trading can produce
a much realistic model of closed market. Such a market shows power
law behavior in terms of wealth (money) distribution within agents
for different situations of money exchange --therefore potentially
advanced.

\end{document}